\documentclass[pre,reprint,showpacs,onecolumn]{revtex4-2}
\usepackage{amssymb,amsmath}
\usepackage{natbib}
\usepackage{graphicx}
\usepackage{color}
\usepackage{bm}
\usepackage{soul}

\newcommand\bra[1]{{\langle{#1}|}}
\newcommand\ket[1]{{|{#1}\rangle}}

\DeclareMathOperator{\Tr}{Tr}

	 \renewcommand{\vec}[1]{\mathbf{#1}}

\graphicspath{{./}{figs/}}

\begin{document}

\title{Quantum Kolmogorov-Sinai entropy and Pesin relation}

\author{Tomer Goldfriend}
\affiliation{Laboratoire  de  Physique  de  l'Ecole  Normale  Sup{\'e}rieure,  ENS,  Universit{\'e}  PSL, CNRS,  Sorbonne  Universit\'e,   Universit\'e Paris-Diderot, Sorbonne Paris Cit\'e,  24  rue  Lhomond,  75005  Paris,  France}

\author{Jorge Kurchan} 
\affiliation{Laboratoire  de  Physique  de  l'Ecole  Normale  Sup{\'e}rieure,  ENS,  Universit{\'e}  PSL, CNRS,  Sorbonne  Universit\'e,   Universit\'e Paris-Diderot, Sorbonne Paris Cit\'e,  24  rue  Lhomond,  75005  Paris,  France}

\date{\today}

\begin{abstract}

We discuss a quantum Kolmogorov-Sinai entropy defined as  the entropy
production per unit time resulting from coupling the system to a weak, auxiliary bath. The expressions we obtain are fully quantum, but require that the system is such that
there is a separation  between the Ehrenfest and the correlation timescales.   We show that they  reduce to
the classical definition in the semiclassical limit, one instance where this separation holds. 
We show a quantum (Pesin) relation between this entropy and the sum of positive eigenvalues of a matrix describing phase-space expansion. Generalizations to the case where entropy grows sublinearly with time are possible. 

\end{abstract}


\maketitle

\section{Introduction}

In Classical Mechanics the relations between thermalisation, ergodicity, entropy and chaos are well defined. The latter is characterized with the Lyapunov exponents and the Kolmogorov-Sinai (KS) entropy of trajectories in phase-space. The analogous situation in closed quantum systems is less understood, e.g., the relation between entanglement entropy and thermalisation~\cite{Kaufman2016}, or entropy production and the quantum Lyapunov exponents. Previous studies expressed the entropy growth rate of a quantum system as  the KS entropy of trajectories evolving in the underlying classical system~\cite{Zurek&Paz1994,Zurek&Paz1995,Pattanayak1999,Miller&Sarkar1999,Monteoliva&Paz2000,Bianchi_etal2015,
Asplund&Berenstein2016}, or which are obtained by projection~\cite{Hallam2019}. Direct quantum  generalizations of the KS entropy were also defined in the past by several rigorous mathematical derivations~\cite{Connes_etal1987,Alicki&Fannes1994}, and more recently, with a simple {\em definition} via a quantity conjectured as playing the role of the quantum  Lyapunov   spectrum,  based on the out-of-time-order-correlator (OTOC)~\cite{Gharibyan2019}. 

In this paper we study a quantum KS entropy by weakly coupling the quantum system to an auxiliary bath--- a set-up which, in the classical case, gives a straightforward heuristic understanding of the equality between entropy production and the sum of positive Lyapunov exponents, the Pesin relation. The idea that in quantum systems, as in classical ones, the sensitivity of a system to an external weak noise can probe chaoticity goes back to Lindblad~\cite{Lindblad1986} (see also~\cite{Zurek&Paz1995,Miller_etal1998}). Here we provide a general and simple definition, which is independent of a specific model and of the existence of a quantum Lyapunov regime. We derive explicitly the semiclassical limit, discuss quantum effects, and show how a quantum Pesin-like relations exist when OTOCs grow exponentially.

The paper is organized as follows: In Sec.~\ref{sec:cl} we present a simple set-up which yields the classical Pesin relation.  The definition for the quantum Kolmogorov-Sinai entropy is introduced in Sec.~\ref{sec:quantum}, { where we obtain expression~\eqref{eq:Gen00}, which is our first general result}. Then, in Sec.~\ref{sec:SClassical} we analyze the  semiclassical limit, and in Sec.~\ref{sec:PesinQ} we derive a quantum Pesin-like relation, relating the KS entropy to the quantum Lyapunov exponents. { This is given in expression~\eqref{eq:Gen4} which is our second result. Its validity is more restricted than Eq.~\eqref{eq:Gen00}, because it assumes exponential growth of some quantities.} In Sec.~\ref{sec:IC} we discuss averaging over initial conditions and how to extend the result to other quantum setups. Finally, outlook and conclusion are given in Sec.~\ref{sec:Dis}.

\section{Classical Pesin relation for the simple-minded}
\label{sec:cl}

\begin{figure}
\centering \includegraphics[angle=0,width=5cm]{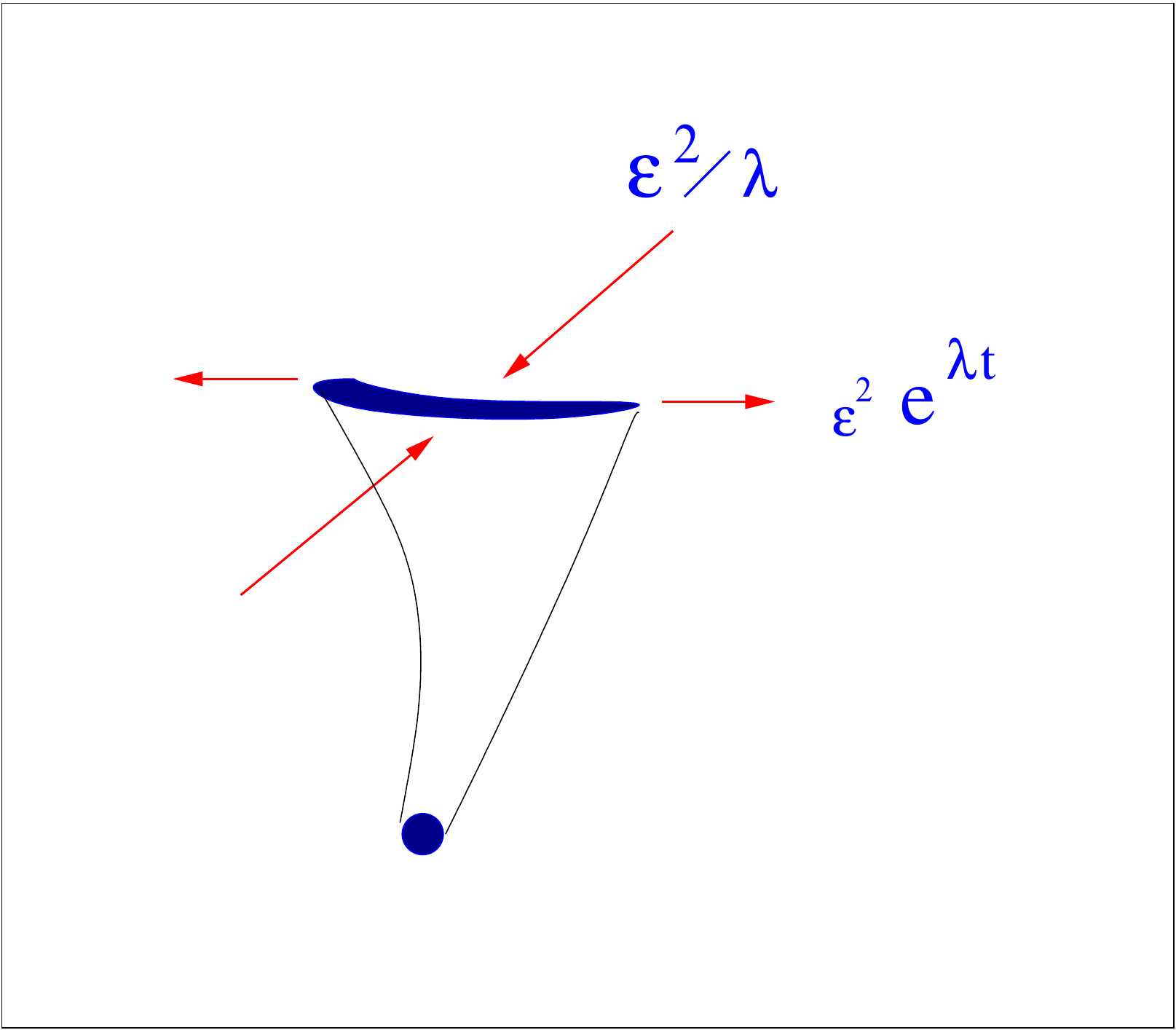}
\caption{The evolution of probability around a trajectory.}
\label{pesin} 
\end{figure}

One way to introduce the Kolmogorov-Sinai entropy is as the entropy production around a trajectory, $\vec{x}\equiv(\vec{q},\vec{p})$, of a system coupled to a weak auxiliary noise $\vec{h}(t)$:
\begin{equation}
\dot x^\alpha(t)=  - \{H,x^\alpha\} + h^\alpha(t),
\end{equation}  
with $H$ being the system's Hamiltonian, $\left\{\cdot,\cdot \right\}$  the usual Poisson brackets, and the index $\alpha$ runs over the components of a $2N-$vector in phase-space.

{ Consider a single classical trajectory.  The Kolmogorov-Sinai entropies $h_{KS}^{(q)}$ (for each $q$) for the perturbed trajectory is defined as \cite{lecomte2007thermodynamic} 
\begin{equation}
t h_{KS}^{(q)}= \frac{1}{1-q}\ln \left\{\int D[{\mbox{history}}] \; P^q[{\mbox{history}}]\right\},
\label{history}
\end{equation}
where the histories are all the trajectories starting from the same point $x(t_0)$ and ending in a point $x^f[h]$, that depends on the noise realization.
This definition may eventually  be supplemented with averaging over initial conditions. We shall be interested in the limit in which the random perturbation is very weak.

Equation (\ref{history}) is not the most easily generalizable to the quantum case, so we shall rewrite it as follows. First, we replicate $a=1,...,q$
\begin{eqnarray}
& & \left\{\int D[{\mbox{history}}] \; P^q[{\mbox{history}}]\right\} =\int_{\bf x_a=x(0)}^{\bf x_a=x^f} \Pi_a D[{\bf x}_a] \int D[{\bf h}] P[{\bf h}] \; \Pi \delta\left(\dot x_a^\alpha(t)+ \{H,x_a^\alpha\} - h^\alpha(t)\right),
\end{eqnarray}
where the delta functions impose the trajectory equations. Then, because these trajectories are casual, we may integrate away all the intermediate coordinates
${\bf x}_a$ except the last, to obtain:
\begin{eqnarray}
& & \int_{\vec x_a=\vec{x}(0)}^{\bf x_a=x^f} \Pi_a D[{\vec x}_a] \int D[{\vec h}] P[{\bf h}] \; \Pi \delta\left(\dot x_a^\alpha(t)+ \{H,x_a^\alpha\} - h^\alpha(t)\right)=\nonumber\\
& & \int \Pi_a d{{\vec x}^f_a}  \int D[{\vec h}] P[{\bf h}] \; \Pi \delta \left({{\vec x}^f_a} -{\vec x}^f[h]\right) =\int \Pi_a d{{\vec x}^f_a}  \int P^q({{\bf x}^f_a}).
\end{eqnarray}
What we have shown is that because of causality, integrals over powers of probabilities of trajectories may be traded for integrals of 
powers of probabilities over the (noise-dependent) endpoint of the trajectories. In other words, we have gone from a description of
`entropy of trajectories' to one of `production of the usual entropy at the end of the trajectory'. 

Starting from  initial configurations within a small ball of radius $a$, and considering  different noise realizations, we have  
at the end of the process at some time $t$ a `probability cloud'. Its distribution  is a function of  the size of the initial set, the noise amplitude, and the stability properties of the trajectory. Denoting the  probability distribution at time $t$  as $P(\vec{x},t)$, we have the R{\'e}nyi entropy of order $q$
\begin{equation}
S_q\equiv\frac{1}{1-q}\ln \left\{\int d {\vec{y}} P^{q} ({\bf y},t)\right\},
\label{eqeq0}
\end{equation}
and the corresponding Kolmogorov-Sinai entropy in terms of its exponential growth: 
\begin{equation}
h_{KS}^{(q)} \equiv  \frac 1 t  [S_q(t) -S_q(t_o)],
\label{eqeq1}
\end{equation}
where $t_o$ is some short reference time. The advantage of this change of point of view is that it is easy to generalize the entropy production
caused by weak noise to the quantum case.
}

Let us now proceed with the classical example.
 Consider the situation as in  Figure (\ref{pesin}): in the absence of noise, 
 in the directions that have exponential expansion between nearby trajectories,  the displacement is $\Delta_{+} \sim a e^{  \lambda^\alpha t}$ 
 with $\lambda^{\alpha}$ being the {\it positive} Lyapunov exponents of the unperturbed system. In an Hamiltonian system,
 there will be exactly as many contracting modes that will compensate the effect of the expanding ones, $\Delta_{-} \sim a e^{ - \lambda^\alpha t}$, and the volume  of the set is unchanged.
 This is just Liouville's theorem, and is a manifestation of the fact that without some form of coarse-graining there is no entropy production in Hamiltonian dynamics.
Let us consider now the effect of the noise: 
{  Roughly, expending and contracting directions are described by $\dot{\Delta}_{\pm}\sim \pm \lambda^{\alpha}\Delta_{\pm} +h(t)$, and the time evolution then gives $\Delta_{\pm}(t)=\int_0^t e^{\pm\lambda^{\alpha}(t-t')}h(t')dt'$. One can see now that the integral which determines $\Delta_+$ is dominated by initial times, $t'\sim 0$, and the displacement induced by the noise  will be amplified  after a time $t$ as  $\Delta_{+} \sim \epsilon e^{  \lambda_\alpha^+ t}$.  However,  now the contracting modes are compensated  by the action of the noise at {\em late times} $t' \sim t$, which
 gives a contribution $ \Delta_{-} \sim \epsilon /\sqrt{\lambda_\alpha^+} $.} 
 Thus, if there are modes exponentially expanding in time, and we are interested in the phase-space volume $V=\Delta_{+}\Delta_{-}$ of the `probability cloud' at the end of the process -- up to logarithmic accuracy
  in time -- we have $\ln V \sim t \sum_{\lambda_i >0} \lambda_i + O(\ln \epsilon)$. This is Pesin's relation.
Let us see this in more detail. 

{ \subsection{Preliminaries} }

{\em i) `Democratic' dimensions} \\

We consider a $2N$-dimensional phase-space ($N$ coordinates and $N$ momenta). We normalize it with the canonical transformation, $\tilde{x}=(b\vec{q},\vec{p}/b)$, with $b$ a parameter with the appropriate dimensions { (and for simplicity from now on we drop the $\tilde{\,}$ notation)}.  Hereafter we take $b=\sqrt{m \omega}$, where $m$ and $\omega$ being mass and frequency of an auxiliary oscillator Hamiltonian that will determine the initial state  (see more details in the semiclassical derivation in Sec.~\ref{sec:SClassical} below). Thus, both position and momentum have units of ${\mbox{\bf (action)}}^{\frac12}$.  \\

{\em ii) White Noise limit}\\

{ The distribution of the noise fields $\vec{h}(t)$ is taken as 
\begin{equation}
 P_{\rm ext}[\vec{h}(t)]\propto e^{-\frac{1}{2\epsilon^2\tau} \sum_\alpha\int dt' h^\alpha(t')^2},
 \label{eq:Pext}
 \end{equation}
where $\tau$ is a typical time-scale for the noise and $\epsilon$ has units of ${\mbox{\bf (energy/time)}}^{\frac12}$. That is, we can think of the random field as taken from a Normal distribution $\mathcal{N}(0,\epsilon^2)$ independently at each time step $\tau$, or a Gaussian  noise of correlation time $\tau$. Assuming that this timescale is shorter than any other relevant one in the system, we have  the `white noise' limit, which  is enough to yield the classical Pesin relation, even if the timescale $\tau$ does not disappear as a parameter of the problem.  We shall assume the shortness of $\tau$  throughout the Paper, also for  the quantum problem.  Note however, that when doubt arise, one has to restore the noise correlations.  In Sec.~\ref{sec:Dis} we discuss how the derivation can be generalized to account for a correlated noise.\\
}

{ \subsection{Entropy of trajectories' end points} }


Deviations from a given trajectory can be described by the Poisson brackets (tangent space), which are given in terms of the second derivative matrix:
{
\begin{equation}
 R^{\alpha \gamma}(t,t')\equiv \frac{\delta x^\alpha(t)}{\delta x^\gamma(t')}= \Omega^{\gamma \beta} \{x^\alpha(t),x^\beta(t')\} 
 =
  \mathcal{T}\exp\left(-\int_{t'}^t dt' \Omega^{\gamma\delta}\{x^\delta(t),\{ x^\alpha(t), {\cal H}\} \}\right), 
\end{equation}%
} 
where we have introduced the $2N\times 2N$ matrix $\Omega = \begin{pmatrix}
0 & 1 \\
-1 & 0
\end{pmatrix}$,  and $\mathcal{T}$ is the time-order operator. { The last equality results from solving the tangent space dynamics for the Poisson brackets~(see Appendix in Ref.\cite{Goldfriend&Kurchan2020}):
$\frac{d}{dt}\{x^\alpha(t),x^\beta(t')\}= \{x^{\gamma}(t),\{x^{\alpha}(t),\mathcal{H}\}\}\Omega^{\gamma\delta}\{x^\delta(t),x^\beta(t')\}$, with the initial condition $\{x^\alpha(t'),x^\beta(t')\}=\Omega^{\alpha\beta}$.
}

The effect of the noise may be computed by considering the deviation it produces from a trajectory { $y^\alpha(t)\equiv x^\alpha(t) - x^{\alpha}_{h=0}(t)$. The tangent space dynamics, to first order in the noise amplitude, then reads
\begin{equation}
y^{\alpha}(t)= \int_0^t dt' R^{\alpha \beta} (t,t')  h^\beta(t').
\end{equation}
}
The $y^\alpha(t)$  are Gaussian variables with zero average, since they are sums of the  $h^\alpha$. { Let us now compute the distribution of the last point $P(\vec{y},t)$, at the end of the process. We start with writing the distribution as a path-integral over realizations of the noise,
$P(\vec{y},t) = \int \Pi_\alpha \frac{\mathcal{D}h^\alpha}{\sqrt{2\pi\epsilon^2{{\tau}}}} \;  \delta\left[\vec{y} - \int_0^t dt' \vec{R} (t,t')  \vec{h}(t')\right] \; e^{-\frac{1}{2\epsilon^2{{\tau}}} \sum_{\alpha}\int dt' h^\alpha(t')^2}$. Then, we Fourier transform the delta-function and solve the Gaussian integral (after completing the square) over the fields $\vec{h}$ to find
\begin{eqnarray}
 P(\vec{y},t) &=&  \int \Pi_\alpha \frac{\mathcal{D}h^\alpha}{\sqrt{2\pi\epsilon^2{{\tau}}}} \; d\hat{{\bf y}} \exp\left\{i\hat{{\bf y}} \left[\vec{y} - \int_0^t dt' \vec{R} (t,t')  \vec{h}(t')\right]  -\frac{1}{2\epsilon^2{{\tau}}} \sum_{\alpha}\int dt' h^\alpha(t')^2\right\} \nonumber
\\
&=& \int  \; d\hat {\bf y} \exp\left\{ i  \vec{y}\hat{\bf{y}}  -\frac{\epsilon^2{{\tau}}}{2} \int_0^t dt' \hat{{\bf y}}^T \vec{R} (t,t') \vec{R}^T (t,t') \hat{{\bf y}}  \right\}.
\end{eqnarray}
The integral over the Fourier variable $\hat{\vec{y}}$ then gives:
\begin{equation}
 P(\vec{y},t) =(2\pi\epsilon^2{{\tau}})^{-d/2} [\; \det {\bf {\cal{A}} }]^{-\frac12}  e^{ -\frac{1}{2 \epsilon^2{{\tau}}}  \vec{y}^T {\cal{A}}^{-1} \vec{y}}, \qquad {\cal{A}}^{\alpha \gamma} (t)\equiv  \; \sum_\beta  \int_0^t dt' R^{\gamma \beta} (t,t') R^{\alpha \beta} (t,t')  = \;  \int_0^t dt' {\cal{B}}^{\alpha \gamma} (t,t').
\label{eq:classical}
\end{equation} 
Eq.~\eqref{eq:classical} thus describes how the weak noise spreads a single trajectory to a probability cloud whose distribution at time $t$ is Gaussian with a variance determined by the $2N\times 2N$ matrix $\cal{A}$--- the main object here.} We note that the matrix ${\bf {\cal{B}}}(t,t')$ is symplectic, it is the instantaneous contribution from time $t'$ to the expansion at time $t$, while
${\bf {\cal{A}}}(t)$ (which is generally not symplectic) is the total expansion. 
{ Taking the final expression for $P(\vec{y},t)$ we can calculate the R{\'e}nyi entropy and its exponential growth rate, according to Eq.~\eqref{eqeq0} and~\eqref{eqeq1} respectively:
\begin{equation}
e^{S_2(t)}=\frac{\mathcal{M}}{2}   \det (\tau \epsilon^2  {\bf  {\cal{A}}(t)} ) ,  \qquad
t h^{(2)}_{KS}= \frac{1}{2}  \ln  \left(    \frac{\det{{\cal{A}}(t)}}{\det{{\cal{A}}(t_o)}} \right),
\label{Renyi1}
\end{equation} where { ${\cal{M}} $ is an irrelevant constant, and $t_o$ is a short reference time.}
}
 For the moment we have not averaged over initial conditions. 

{ \subsection{Pesin relation} 

In the case in which the second derivative matrix $\{x^{\delta},\{x^{\alpha},\mathcal{H}\}\}$ is constant in time, one can easily confirm the claim above: if there is an exponential expansion in a mode of ${\bf {\cal{B}}}$, then the initial times $t' \sim 0$  dominate the time-integral and  the expanding directions are of order $\epsilon e^{ \lambda_\alpha t}$. Contracting modes of ${\bf {\cal{B}}}$ are actually stopped from contracting completely by the action of the noise at {\em late times} $t' \sim t$, which then  give a contribution $ \epsilon /\sqrt{\lambda_\beta} $.
The general case is similar but  slightly less trivial, as the tangent space vectors rotate in time. A rigorous approach is given by defining covariant Lyapunov vectors, as we explain now  (a different derivation is also given in  Appendix~\ref{app:susy}). 

The entropy production in Eq.~\eqref{Renyi1} depends on the determinant of the matrix $\mathcal{A}$. First, we use the facts that $\vec{R}(t,t')=\vec{R}(t,0)\vec{R}(0,t')$ and $\vec{R}(0,t')=\vec{R}^{-1}(t',0)$, to write
\begin{equation}
\left| {\cal{A}} (t)\right| =\left| \int_0^t dt' \vec{R}(t,t') \vec{R}^T (t,t') \right|
=\left| \int_0^t dt' \vec{R}(t,0)\vec{R}^{-1}(t',0) \left(\vec{R}^T(t',0)\right)^{-1}\vec{R}^T(t,0) \right|.
\label{eq:cl1}
\end{equation}
Then, symplecticity of $\vec{R}$, namely $\vec{R}^{-1}= \Omega^{-1}\vec{R}^T\Omega$, gives us 
\begin{multline}
\left| {\cal{A}} (t)\right| =\left| \int_0^t dt' \vec{R}(t,0)\Omega^{-1}\vec{R}^{T}(t',0)\vec{R}(t',0)\Omega\vec{R}^T(t,0) \right|=\left| \vec{R}(t,0)\Omega^{-1}\right|\left| \int_0^t dt' \vec{R}^{T}(t',0)\vec{R}(t',0)\right|\left|\Omega\vec{R}^T(t,0) \right|=\\ \left| \vec{R}(t,0)\Omega^{-1}\vec{R}^T(t,0)\Omega \right|\left| \int_0^t dt' \vec{R}^{T}(t',0)\vec{R}(t',0)\right|= \left| \int_0^t dt' \vec{R}^{T}(t',0)\vec{R}(t',0)\right|.
\label{eq:cl2}
\end{multline}
As a final step, we use the notion of covariant Lyapunov vectors~\cite{CLV1,CLV2,CLV3}: for any time $t_1$ we have $2N$ vectors~\footnote{For simplicity we assume that there is no degeneracy and all Lyapunov exponents are distinct.} $\gamma_{k}(t_1)$ such that $||\vec{R}(t'+t_1,t_1)\gamma_{k}(t_1)||\rightarrow \exp(\lambda_k t')$ as $t'\rightarrow \infty$, with $\lambda_k$ being  the Lyapunov exponent. Let us define the $2N \times 2N$ matrix $\Gamma(0)$ whose columns are the covariant Lyapunov vectors at time $t=0$ and write
\begin{equation}
\left| {\cal{A}} (t)\right|=\left| \left[\Gamma(0)\Gamma^T(0)\right]^{-1}\right| \left| \Gamma^T (0) {\cal{A}} (t) \Gamma(0)\right|=\left| \left[\Gamma(0)\Gamma^T(0)\right]^{-1}\right| \left|  \int_0^t dt' \Gamma^T (0)\vec{R}^{T}(t',0)\vec{R}(t',0) \Gamma(0)\right|. 
\end{equation}
Therefore, at times $t'$ large enough within the integral we have a matrix of the form ${\cal{B}}^{\beta\alpha} (t',0)\sim  v^{\alpha}(t')v^{\beta}(t') e^{(\lambda_\alpha+\lambda_\beta) t'}$, leading to  $\left| {\cal{A}} (t)\right|\rightarrow e^{\sum _{\lambda_\alpha>0}\lambda_\alpha t}$ at $t\rightarrow \infty$.

}
 
{ \subsection{Anticipating the quantum case} }

 { We now go back to the classical result in Eq.~\eqref{Renyi1}. The left hand side of this equation is also valid even in  the absence of exponential divergences, although in that case the total entropy production does not scale linearly with time, and the KS entropy defined as a logarithmic entropy production {\em per unit time} goes to zero.    This is also true for the quantum formula derived below. Two additional comments will be important for the quantum case: \\

{\em a) Coupling to the noise:} \\

Is it necessary to couple noise to all variables? The question is relevant classically because we usually couple noise only to the coordinates, and it appears in half of
 Hamilton's equations. It will be much more serious quantum mechanically, where there is considerably more freedom.
Let us see the case when  the noise is coupled to a single variable, say $x^1$.  We may redo the calculation above, to obtain a new matrix ${\bf \hat {\cal{A}}}$ given by
\begin{equation}
\hat {\cal{A}}^{\alpha \beta} \equiv \int dt'  \; R^{\alpha,1}(t,t') R^{\beta,1}(t,t').
\end{equation}
 Once again, if we consider only the initial time $t'=0$, the corresponding matrix will have rank one, and we shall obtain only one expansion direction.
 However, as we integrate over many $t'$, {\em provided the system is such that the eigenbasis is rotating with dynamics}, we have that the rank of the matrix 
 will increase as time passes. This applies to the `recent' times as well, which also have to be integrated so that they contribute fully to the contracting directions.
 We hence conclude that if at times $t$ for which the KS entropy is computed, the tangent space has rotated ergodically, then it is not important how many
 noises we have coupled to the system.
 { We shall see below that quantum mechanically the situation is more subtle, and also more unavoidable.}  
 Let us note also that applying noise to a local (in space) operator, and seeing how entropy builds up, is an interesting question in itself.\\

{\em b)  Initial distribution width.}\\

Suppose the initial condition $\vec{y}_0$, rather than being a single configuration, is spread over an isotropic Gaussian of width $a$ in phase-space, as in Figure \ref{pesin}. 
{ A simple modification of the steps  leading to Eq.~\eqref{eq:classical},  namely, taking $\vec{y}= \int_0^t dt' \vec{R}(t,t')\vec{h}(t') + \vec{R}(t,0)\vec{y}_0$,  gives 
\begin{equation}
S_{2} (t) -S_2(t=0) =\frac{1}{2}  \ln \det \left[ {\epsilon^2\tau} {\cal{A}}(t)+ {a^2} \; { {\cal{B}}}(t,0) \right] + {\mbox{const}}.
\label{Renyi2}
\end{equation}
In the absence of noise $\epsilon^2{\cal{A}}(t)=0$ and  the rhs is a constant, because ${\cal{B}}$ is symplectic and has determinant one. Note that $a^2$, being the volume of a $(p,q)$ cell, has dimensions of action.
We obtain the correct time-dependence  {\em only if the expansion due to noise  is large compared
to the initial volume:  $\frac{\epsilon^2 \tau}{a^2}  e^{\lambda t} \gg 1$}}. 
This gives us a warning of trouble ahead for the quantum case,
where initial packet width is unavoidable, and we are not allowed to take times to infinity.\\~

\section{Quantum Kolmogorov-Sinai entropy: a definition}
\label{sec:quantum}

Motivated by the classical case, we consider a quantum evolution  coupling the system to a weak, auxiliary  bath, which can be introduced from first principles as a coupling to an infinite set of oscillators, exactly the way one derives the classical Langevin equation~\cite{Zwanzig1973}.  
We note that this line of reasoning was applied for understanding entropy production and decoherence in open quantum systems~\cite{Zurek&Paz1994,Monteoliva&Paz2000,Pattanayak1999}. 

If the temperature of the bath is  infinite, friction may be neglected, otherwise  forward and backward propagations are with different fluctuating fields, as may be seen from the Schwinger-Keldysh formalism~\cite{Kamenev2011,Cugliandolo&Lozano1999}. In this limit the bath becomes effectively classical.
We can thus consider, for infinite auxiliary bath temperature,  the Hamiltonian $H=H_{0}(\vec{x})+\sum_{\alpha}h^\alpha(t)x^{\alpha}(t)$, with the  time-dependent random field $\vec{h}(t)$, which has probability distribution $P_{\rm ext}[\vec{h}(t)]$ given in Eq.~\eqref{eq:Pext}. For simplicity, as in the classical case, we shall consider that the correlation time of this bath $\tau$ is much smaller than the relevant timescales of the problem at hand (see discussion above concerning  the white noise limit).

We are interested first in the  R{\'e}nyi entropy generated in the process, starting from an initial density matrix $\rho_0$. For the moment we restrict ourselves for $q=2$, in Appendix~\ref{app:Renyi} we generalize for all integer $q$. { The  expression for $S_2(t)$ involves path-integral over two different realizations of the noise, $\vec{h}_1$ and $\vec{h}_2$:}
\begin{equation}
\Tr \left\{\rho^2(t)\right\}= 
\Tr \left\{\int\mathcal{D}\vec{h_1}\mathcal{D}\vec{h_2}P[\vec{h_1}]P[\vec{h_2}]U^{\dagger}_{h_2,I}U_{h_1,I}\rho_{0}U^{\dagger}_{h_1,I}U_{h_2,I}\rho_{0}\right\},
\label{eq:Trho2}
\end{equation}
where we work with the interaction picture $U_{h,I}=\mathcal{T}\exp\left( -\frac{i}{\hbar}\int\sum_{\alpha} h^{\alpha}(t')x^{\alpha}(t')dt' \right)$, with the operator $x(t)$  given in the {\em Heisenberg picture of the unperturbed dynamics}, $\vec{x}(t)=e^{iH_0 t/\hbar}\vec{x}e^{-iH_0 t /\hbar}$.  

 Our approach to evaluate Eq.~\eqref{eq:Trho2} in the weak noise limit is as follows: we first approximate $\ln \Tr\rho^2(t)$ for a given random field, up to second order in $\vec{h}_{1,2}$, and then we average over the noise amplitude. The same procedure can be carried out for the analogous classical problem, which then reproduces the classical result in Eq.~\eqref{Renyi2}; this is shown explicitly in Appendix~\ref{app:Poisson}. 
Let us then define
\begin{equation}
\mathcal{Z}[\vec{h_1},\vec{h_2}]\equiv \ln\Tr \left\{U^{\dagger}_{h_2,I}U_{h_1,I}\rho_{0}U^{\dagger}_{h_1,I}U_{h_2,I}\rho_{0}\right\}, \qquad { \Tr \left\{\rho^2(t)\right\}= \int\mathcal{D}\vec{h_1}\mathcal{D}\vec{h_2}P[\vec{h_1}]P[\vec{h_2}]e^{\mathcal{Z}[\vec{h_1},\vec{h_2}]},}
\end{equation}
and perform time-dependent perturbation theory  around $\vec{h}_1,\vec{h}_2=0$, up to second order. The first order terms vanish, whereas the second order terms depend only on the difference ${\bm \eta}=\vec{h}_1-\vec{h}_2$, we find (see derivation in Appendix~\ref{app:Taylor}):
\begin{equation}
\Tr \rho^2(t)\approx 
\Tr\rho_0^2\int\mathcal{D}{\bm \eta}P_{\rm ext}[{\bm \eta}]
\exp\left(
-\frac{1}{2}\sum_{\alpha,\beta}\int dt'\,dt''\, \right. \\ \left.  \frac{\Tr\left\{i[x^{\alpha}(t'),\rho_{0}]i[x^{\beta}(t''),\rho_{0}]\right\}}{\hbar^2\Tr\rho^2_0} \eta^{\alpha}(t') \eta^{\beta}(t'')\right).
\label{eq:Gen0}
\end{equation}
 We are left with solving Eq.~\eqref{eq:Gen0} as a Gaussian integral over the fields $\vec{\eta}^{\alpha}(t)$, recalling that, in the limit
 of small timescale $\tau$ of the bath: $P_{\rm ext}[{\bm \eta}]\propto e^{-\frac{1}{2\epsilon^2\tau}\sum_{\alpha}\int dt'(\eta^{\alpha}(t'))^2}$. Next, we obtain this solution by taking a simplified initial state $\rho_0$, and employing { the Matrix Determinant Lemma.}

{ For simplicity we work hereafter with a Hilbert space which is represented by a real, orthogonal basis $\{\ket{\mu}\}$. Let us consider the case  where the initial density matrix is given by a pure state, $\rho_0=\ket{\phi_0}\bra{\phi_0}$. We shall generalize in Sec.~\ref{sec:IC} to any initial condition.} 
Developing the trace in the exponent of Eq.~\eqref{eq:Gen0} we can write:
\begin{equation}
\Tr\left\{[x^{\alpha}(t'),\rho_{0}][x^{\beta}(t''),\rho_{0}]\right\}=\bra{\phi_0} [x_c^{\alpha}(t')x_c^{\beta}(t'')  + x_c^{\beta}(t'')x_c^{\alpha}(t')] \ket{\phi_0},
\label{commu}
\end{equation}
where
\begin{equation}
x_c^{\alpha}(t)\equiv x^{\alpha}(t)-\bra{\phi_0}x^{\alpha}(t)\ket{\phi_0}.
\end{equation}
We thus find: 
\begin{eqnarray}
\Tr \rho^2(t) &\approx& 
\int\mathcal{D}{\bm \eta}P[{\bm \eta}]
e^{
-\frac{1}{\hbar^2}\sum_{\mu}\left|\sum_{\alpha}\int_0^t dt'\, \bra{\phi_0}x_c^{\alpha}(t')\ket{\mu} \eta^{\alpha}(t')\right|^2}= \nonumber\\
&=&
\int\mathcal{D}{\bm \eta}P[{\bm \eta}]
e^{
-\frac{1}{\hbar^2}\sum_{\mu}\left( \left\{\sum_{\alpha}\int_0^t dt'\, \bra{\phi_0}x_c^{\alpha}(t')\ket{\mu}_{\rm Re} \eta^{\alpha}(t')\right\}^2+ 
\left\{\sum_{\alpha}\int_0^t dt'\, \bra{\phi_0}x_c^{\alpha}(t')\ket{\mu}_{\rm Im} \eta^{\alpha}(t')\right\}^2\right)},
\label{eq:Gen01}
\end{eqnarray}
where we { recall} that the wavefunctions $\ket{\mu}$ (and the noise) are real, { and the subscripts ${\rm Re}$ and ${\rm Im}$ refer, respectively, to the real and imaginary parts of a complex number, e.g.,  $\bra{\phi_0}x_c^{\alpha}(t')\ket{\mu}_{\rm Re}\equiv {\rm Re}\left( \bra{\phi_0}x_c^{\alpha}(t')\ket{\mu} \right)$. We now simplify the solution to the Gaussian integral in Eq.~\eqref{eq:Gen01} by using the Matrix Determinant Lemma: for an invertible matrix $\vec{K}$ of order $k\times k$, and a matrix $\vec{U}$ of order $k\times m$ we have:
\begin{equation}
\det\left(\vec{K}+\vec{U}\vec{U}^T\right)=\det\left(\vec{I}_m+\vec{U}^T\vec{K}^{-1}\vec{U}\right)\det(\vec{K}),
\label{eq:MDL}
\end{equation} 
where $\vec{I}_m$ is the $m\times m$ identity matrix. If we now discretize time, $0=t_1<t_2<\dots<t_T=t$, we can rewrite the path-integral as an integral over a vector $\eta^{\alpha}_{t_n}\equiv \eta^{\alpha}(t_n)$ of length $2NT$
\begin{equation}
\Tr \rho^2(t) = \det\left|\frac{\vec{K}}{\epsilon^2\tau}\right|^{1/2}
\int d{\bm \eta} \exp\left[-\frac{1}{2\epsilon^2\tau} \eta^{\alpha}_{t_n} K^{\alpha\beta}_{t_nt_m} \eta^{\beta}_{t_m}
-\frac{1}{2\hbar^2}\eta^{\alpha}_{t_n} \left( U^{\alpha;\mu}_{t_n} U^{\beta;\mu}_{t_m} +V^{\alpha;\mu}_{t_n} V^{\beta;\mu}_{t_m} \right)\eta^{\beta}_{t_m})\right],
\label{eq:UV}
\end{equation} 
where we define the $2NT\times 2NT$ matrix $K^{\alpha\beta}_{t_nt_m}\equiv  \delta^{\alpha\beta}_{t_nt_m} dt$ and the $2NT\times  L$ matrices $U^{\alpha;\mu}_{t_n}\equiv 2\bra{\phi_0}x_c^{\alpha}(t_n)\ket{\mu}_{\rm Re}dt$ and $V^{\alpha;\mu}_{t_n}\equiv 2\bra{\phi_0}x_c^{\alpha}(t_n)\ket{\mu}_{\rm Im}dt$, with  $L$ being the size of the Hilbert space and $N$ the number of degrees of freedom. The solution to the Gaussian integral involves a determinant of a $2NT\times 2NT$ matrix, which, by employing the Matrix Determinant Lemma can be simplified to a determinant of a $2L\times 2L$ matrix:
\begin{equation}
\Tr \rho^2(t) = \det\left|\frac{\vec{K}}{\epsilon^2\tau}\right|^{1/2}\det\left|\frac{\vec{K}}{\epsilon^2\tau}+
\frac{1}{\hbar^2}\begin{pmatrix}
\vec{U} & \vec{V}
\end{pmatrix}
\begin{pmatrix}
\vec{U}^T\\
\vec{V}^T
\end{pmatrix}
\right|^{-1/2}=\det\left|\vec{I}_{2L}+
\frac{\epsilon^2\tau}{\hbar^2}
\begin{pmatrix}
\vec{U}^T\\
\vec{V}^T
\end{pmatrix}\vec{K}^{-1}\begin{pmatrix}
\vec{U} & \vec{V}
\end{pmatrix}
\right|^{-1/2}.
\end{equation}
Inserting back the definitions of $\vec{K}$, $\vec{U}$, and $\vec{V}$, with taking back the continuous time limit we find
\begin{equation}
\Tr \rho^2(t) = \det\left[1+2\frac{\epsilon^2\tau}{\hbar^2}\sum_{ \alpha }\int_0^tdt'\,
\begin{pmatrix}
\left(\bra{\phi_0}x_c^{\alpha}(t')\right)_{\rm Re} \\
\left(\bra{\phi_0}x_c^{\alpha}(t')\right)_{\rm Im}
\end{pmatrix}
\begin{pmatrix}
\left(\bra{\phi_0}x_c^{\alpha}(t')\right)^T_{\rm Re} 
& \left(\bra{\phi_0}x_c^{\alpha}(t')\right)^T_{\rm Im}
\end{pmatrix}\right]^{-1/2}.
\end{equation}
}
Note that   the determinant is of a matrix that:  {\em i)} has, at this stage, twice the dimension of  Hilbert space $2L\times 2L$,  {\em ii) } because it is a sum over tensor products it has positive eigenvalues. As a final step we may multiply the matrix within the determinant left and right with { $\vec{Z}\equiv\frac{1}{\sqrt{2}}\begin{pmatrix}\vec{I}_{L} & i\vec{I}_{L} \\ \vec{I}_{L} &-i\vec{I}_{L}\end{pmatrix}$ and $\vec{Z}^{-1}$ respectively to find:
{\begin{equation}\boxed{
\Tr \rho^2(t)  =\det\left[1+\frac{\epsilon^2\tau}{\hbar^2}{\bf A}\right]^{-1/2} = e^{-[S_{(2)}(t) -S_{(2)}(0)] } =  e^{-t h_{KS}^{(2)} }
} 
\label{eq:Gen00}  
\end{equation} 
}
\begin{equation}
{\bf A} \equiv  \sum_{ \alpha }\int_0^tdt'\,
\begin{pmatrix}
x_c^{\alpha}(t')\ket{\phi_0}  \bra{\phi_0}    x_c^{\alpha}(t')   \qquad   x_c^{\alpha}(t')\ket{\phi_0}   \bra{\phi_0}  x_c^{\alpha*}(t') \\
x_c^{\alpha*}(t')\ket{\phi_0}  \bra{\phi_0} x_c^{\alpha}(t')   \qquad      x_c^{\alpha*}(t')\ket{\phi_0}  \bra{\phi_0} x_c^{\alpha*}(t')
\end{pmatrix}= \sum_{ \alpha }\int_0^tdt'\,
\begin{pmatrix}
x_c^{\alpha}(t')\ket{\phi_0}   \\
x_c^{\alpha*}(t')\ket{\phi_0}
\end{pmatrix}
\otimes
\begin{pmatrix}
\bra{\phi_0} x_c^{\alpha}(t')  \\
\bra{\phi_0} x_c^{\alpha*}(t')
\end{pmatrix},
\label{eq:Gen1}
\end{equation}
}
where we still assume that $\ket{\phi_0}$ and the Hilbert space bases are  real.

Equation~\eqref{eq:Gen00} with { definition (\ref{eq:Gen1})} is our first main result: a formula for the entropy production starting from $\ket{\phi_0}\bra{\phi_0} $
due to a weak coupling to a bath. It is valid even when there is no exponential expansion, but if and when there is,  we deduce a  generalization to the classical Kolmogorov-Sinai entropy.  
In Appendix \ref{app:Renyi} we repeat the calculation for a general R{\'e}nyi entropy: the result is that these entropies differ when averaged over different starting situations (`multifractality', just as in the classical case), but the effect of quantum fluctuations alone is subdominant. There is no multifractality of quantum origin for a single trajectory.

A general comment is in order here: 
The fluctuating field is coupled only to the phase-space operators $x^{\alpha}$. However, as discussed at the end of Sec.~\ref{sec:cl}, this can be generalized easily to any coupling:  if we consider a set of observables $\{\mathcal{O}^{\alpha}\}$, and the Hamiltonian $H(\vec{x})=H_0(\vec{x})+\sum_{\alpha}h^{\alpha}(t)\mathcal{O}^{\alpha}$, then   the expression
(\ref{eq:Gen1}) remains the same, with the $\{x_c^\alpha\}$ replaced by $\{\mathcal{O}_c^{\alpha}\}$. So far this situation is very much like the classical case, but
{\em there is however a catch}:  quantum mechanically we may in principle couple noises to operators that are not simple functions of the physical operators $x_i$,
for example the projector onto any specific wavefunction, so one could have an exponential number of independent noises applied to the system. We are not choosing to do this here, 
the noise terms are acting on physical observables, and are in number of the same order as the  degrees of freedom. This will have  the important consequence that some final expressions
may be written in lower-dimensional spaces.

For the rest of the paper we further discuss our quantum entropy production Eq.~\eqref{eq:Gen00}, and the KS entropy that it defines in the presence of an exponential growth.

\subsection{Additivity}  
\label{sec:add}

The KS entropy is a rate of entropy production. As such, our definition in Eq.~\eqref{eq:Gen00} must admit the property of additivity. To verify this,  let us consider two uncoupled systems in the Hilbert space $\mathcal{H}^{(1)}\otimes\mathcal{H}^{(2)}$ of size $L^2$. The matrix ${\bf A}$ in Eq.~\eqref{eq:Gen1} is now constituted of terms of the form:
\begin{equation}
z_c^{(1)} \left[\ket{\phi_0^{(1)}}\otimes  \ket{\phi_0^{(2)}} \right] \; \left[\bra{\phi_0^{(1)}} \otimes \bra{\phi_0^{(2)} } \right] \bar z_c^{(1)} \qquad {\mbox{and}} \qquad z_c^{(2)} \left[\ket{\phi_0^{(1)}}\otimes  \ket{\phi_0^{(2)}}  \right] \;\left[\bra{\phi_0^{(1)}} \otimes \bra{\phi_0^{(2)} } \right] \bar z_c^{(2)},
\end{equation}
where the superindices denote the space to which each $z$ belongs.
These may be written as
\begin{equation}
 \ket{\phi_0^{(2)}} \bra{\phi_0^{(2)} } \;\;  \otimes \;\; z_c^{(1)}\ket{\phi_0^{(1)}}  \bra{\phi_0^{(1)}}  \bar z_c^{(1)} \qquad {\mbox{and}} \qquad  \ket{\phi_0^{(1)}}\bra{\phi_0^{(1)}}\;\; \otimes  \;\; z_c^{(2)}\ket{\phi_0^{(2)}}   \bra{\phi_0^{(2)} } 
\bar z_c^{(2)}. 
 \label{produ}
 \end{equation}
Consider now a basis for each space $\{\ket{\phi_0}, \ket{\mu}\}$ where the   $\ket{\mu}$ are orthogonal to $\ket{\phi_0}$. In the tensor product of the two bases we have vectors like
{\it i)} $\ket{\phi_0}\otimes\ket{\phi_0}\,,\,{\it ii) } \ket{\phi_0}\otimes\ket{\mu}\,,\,{\it iii) } \ket{\mu}\otimes\ket{\phi_0}\,,\,{\it iv) } \ket{\mu}\otimes\ket{\mu}$. Within subspaces {\em i} and  {\em iv} all terms are zero. The l.h.s. terms of Eq.~(\ref{produ}) are zero in subspace {\em ii}   and equal to $z_c^{(1)}\ket{\phi_0^{(1)}}  \bra{\phi_0^{(1)}}  \bar z_c^{(1)}$ in subspace {\em iii}. Similarly, the r.h.s. terms of Eq.~(\ref{produ}) are only
non-zero in subspace {\em ii}, where they take the value $z_c^{(2)}\ket{\phi_0^{(2)}}   \bra{\phi_0^{(2)} }\bar z_c^{(2)}$. The determinant of Eq.~\eqref{eq:Gen1} is the determinant of a matrix with two different
blocks, each of size $2L-1\times 2L-1$, and the identity elsewhere. The determinant is then a product of the ones associated with each space, so that the logarithm is additive.

\subsection{Timescales}
\label{sec:interp}

We will check below in Sec.~\ref{sec:SClassical} that our definition of the KS entropy gives the correct result in the semiclassical limit. However, for a truly quantum system, Eq.~(\ref{eq:Gen1}) seems problematic:
if we let $\epsilon \rightarrow 0$ at finite $\hbar$, we will not get a finite limit for $h_{KS}$ plus logarithmic corrections in $\epsilon$, but rather 
 $ \Tr \left\{\rho^2(t)\right\} \sim_{\epsilon \rightarrow 0} \frac{\epsilon^2\tau}{\hbar^2} \Tr A$. And yet, as we have seen when discussing additivity, the  identity term within the determinant   is essential to obtain a meaningful result,  so we ought to understand what is its meaning. We have a dilemma: we need $\epsilon$ to be very small  because we have assumed this in all our expansions, but we now see that in the
 limit $\epsilon \rightarrow 0$ the expression becomes meaningless. There is an apparently obvious way out to this: choose times that are long enough, so that the exponential expansion of ${\bf A}$ compensates
 for the smallness of $\epsilon$. However, we expect there will be a limit to this, the Ehrenfest time at which a minimal quantum packet will expand throughout phase-space.
 {\em We argue that this is an inherent problem, shared also by the very definition of quantum  Lyapunov exponents themselves. }

  Let us  take the dimensional  factors away from the phase-space coordinates (which we had normalized to
 have dimensions of (action)$^{\frac 12}$), writing  $x'= \sqrt{\frac {1}{\hbar}}x$
and  define ${\bf A_o}$ as in Eq.~\eqref{eq:Gen1} but with $x'$, to get: 
\begin{equation}
e^{-h^{(2)}_{KS} t } \sim \Tr \left\{\rho^2(t)\right\}=\det \left|1+\frac{\epsilon^2\tau}{\hbar}A_o \right|^{-1/2}.
\label{determ}
 \end{equation}
We recognize that the second term has the interpretation of a phase-space expansion {\em measured as the number of `quantum cell' sizes}.  Thus, the comparison with the first term is just expressing the fact that 
expansion of a direction over a fraction of one quantum cell size does not contribute to the determinant.

 Let us assume that there is, at least during some time, a Lyapunov expansion  with a largest exponent $\lambda_1$, meaning that  ${\bf A_o}$ will have some elements growing as $\frac{1}{\lambda_1} e^{\lambda_1 t}$.
This expansion may last up to the Ehrenfest/scrambling time $t_E$, which we may estimate as:
\begin{equation}
e^{\lambda_1\; t_E} \sim \; \frac k \hbar ,
\end{equation}
where $k$ depends on the system, and in particular its number of degrees of freedom. { (We note that its value is rarely specified in the literature -- we shall not break with this tradition, but just specify that it has to be related to the unperturbed system, it does {\em not} have to involve $\epsilon^2 \tau$)}. 
Looking at (\ref{determ}), we conclude that we may hope to have a meaningful expression, valid at times just below the Ehrenfest time, if:
\begin{equation}
\frac{\epsilon^2\tau}{\lambda_1 \hbar^2} k \gg 1.
\end{equation}
It is now clear that the semiclassical limit, taken {\em before} $\epsilon \rightarrow 0$, is a possibility to have a well-defined KS entropy. 
But there is more: there are cases when the constant $k$ is expected to scale as the number of degrees of freedom $k \sim N$, while the ordinary two-point correlation does not.
For these kind of systems, where there is an hierarchy between Ehrenfest/scrambling and correlation times (the reader will find a discussion of this in Ref.~\cite{Maldacena_etal2016}), we may take $\epsilon \rightarrow 0$ with  
${\epsilon^2 \tau}{k}$ large, and the Kolmogorov-Sinai entropy is well defined,  and  a non-zero value for it is {\em possible}.

\section{semiclassical}
\label{sec:SClassical}

Let us now evaluate our quantum definition for the KS entropy in the semiclassical limit. This can be done starting with the earlier expression Eq.~\eqref{eq:Gen0}, as we show in Appendix~\ref{app:sc2}; or, working with Eq.~\eqref{eq:Gen1}, as we show now. We need to evaluate the expectation value $\bra{\phi_0}x_c^{\alpha}(t')\ket{\mu}$ in the semiclassical limit, where, for simplicity  of presentation we assume that the system evolves with the unperturbed Hamiltonian of the form $H_0(\vec{x})=\frac{\vec{p}^2}{2m}+V(\vec{q})$. The unitary evolution in the semiclassical limit can be obtained in a standard way, developing the path-integral  around the classical trajectory: we have $\vec{x}=\vec{x}_{\rm cl}(t)+\vec{y}(t)$, with $\vec{x}_{\rm cl}(t)$ being the classical trajectory, and its quantum perturbation is described by the operator $\vec{y}=(\vec{\delta p},\vec{\delta q})$ through the time-dependent Hamiltonian
\begin{equation}
H_{\rm sc}(\vec{y})= \vec{y}^T 
\begin{pmatrix}
\frac{1}{2m}  & 0 \\
0 & \frac{V''(t)}{2}
\end{pmatrix}
\vec{y},\qquad V''(t)\equiv \left.\frac{\partial^2 V }{\partial q^{\alpha} \partial q^{\beta}}\right|_{x_{\rm cl}(t)}.
\end{equation}
This is an harmonic oscillator with time-dependent frequencies: the evolution of $\vec{y}$ is linear and it coincides with the classical counterpart. Therefore, $y^\alpha(t)=R^{\alpha\beta}(t,0) y^{\beta}_0$, with
\begin{equation}
\vec{R}(t,0)\equiv \mathcal{T}\exp\left[\int^t_0 dt' \begin{pmatrix}
0 & \frac{1}{m} \\
-V''({\bf x_{cl}}(t)) & 0
\end{pmatrix}
\right].
\label{yyy}
\end{equation} 
Since, by construction, $\vec{x}_{\rm cl}$ is diagonal and $\bra{\phi_0}\vec{y}\ket{\phi_0}=0$, we have $\bra{\phi_0}x^{\alpha}_c(t')\ket{\mu}=\bra{\phi_0}y^{\alpha}(t')\ket{\mu}$.  
The vectors which build the operator in Eq.~(\ref{eq:Gen1}) then read
\begin{equation}
\bra{\phi_0}x^{\alpha}_c(t')=R^{\alpha\beta}(t',0)\bra{\phi_0}y^{\beta}(0).
\label{eq:sc_element}
\end{equation}

{ 
We now choose to work with a specific basis: the one of a set of $N$ harmonic oscillators of mass $m$ and frequency $\omega$. As described in the classical case, we normalize the operators such that they have the same dimensions, {\bf (action)}$^{\frac 12}$, taking $x=(b\vec{q},\vec{p}/b)$ with $b=\sqrt{m\omega}$, which we shall assume in what follows.  The oscillator Hamiltonian thus reads: $H_{\rm ref}=\frac{\omega}{2}\sum_a  \left[ (p^a)^2 +  (q^a)^2 \right]$.  We take the ground-state of this  Hamiltonian to define the initial density $\ket{\phi_0}=\ket{0}$. This choice of basis and initial conditions is appropriate to obtain the semiclassical limit since the initial distribution is localized in phase-space. 

Let us now calculate the elements in Eq.~\eqref{eq:sc_element} which gives the matrix $\vec{A}$ in the semiclassical limit. The Hilbert space  breaks into subspaces of different total numbers of boson excitations of the oscillators Hamiltonian $H_{\rm ref}$. The $y^\alpha(0)$, being linear,
 create,  when acting on $\ket{\phi_0}$, states of one boson. 
Hence, in the space having two or more bosons the matrix  $\vec{A}$ is zero, and the determinant in Eq.~\eqref{eq:Gen00} is determined by a $2N$-dimensional matrix within the doubled one-boson subspace (recall that the dimensions of $\bf A$ is twice the size of the Hilbert space). This matrix is constructed by the $2N \times N$ blocks $W^{\alpha n}\equiv R^{\alpha\beta}(t',0)\bra{\phi_0}y^{\beta}(0)\ket{1^{n}}=
\sqrt{\frac{\hbar}{2}} \left( R^{\alpha q^n}(t',0)-iR^{\alpha p^n}(t',0)\right)$,
as
\begin{equation}
 \left.\vec{A}\right|_{\left\{\ket{1}^N,\ket{1}^N\right\}} \xrightarrow[\hbar\to 0]{}
\begin{pmatrix}
\vec{W}^T\vec{W} & \vec{W}^T\vec{W}^* \\
\vec{W}^{*T}\vec{W}& \vec{W}^{*T}\vec{W}^*
\end{pmatrix},
\end{equation}
where $\ket{1^n}$ is the state of one boson for the $n-$th degree of freedom and zero for all the others. Note the crucial role played by the identity in  Eq.~(\ref{eq:Gen00}) outside this subspace.  Since we are interested in the determinant we can multiply the above matrix from left and right with $\tilde{\vec{Z}}^{-1}\equiv\frac{1}{\sqrt{2}}\begin{pmatrix}\vec{I}_{L} & \vec{I}_{L} \\ i\vec{I}_{L} &-i\vec{I}_{L}\end{pmatrix}$ and $\tilde{\vec{Z}}$ respectively, to find  
\begin{equation}
 \left.A^{\alpha\gamma}\right|_{\left\{\ket{1}^N,\ket{1}^N\right\}} \xrightarrow[\hbar\to 0]{}\sum_\alpha\int dt\, 
\begin{pmatrix}
R^{\alpha q^\beta}(t',0)R^{\alpha q^\gamma}(t',0) & R^{\alpha p^\beta}(t',0)R^{\alpha p^\gamma}(t',0) \\
R^{\alpha p^\beta}(t',0)R^{\alpha q^\gamma}(t',0)& R^{\alpha p^\beta}(t',0)R^{\alpha p^\gamma}(t',0)
\end{pmatrix}.
\end{equation}
}
We thus show that in the semiclassical limit the second Renyi entropy is given by
\begin{equation}
\ln \Tr \left\{\rho^2(t)\right\}=\frac{1}{2}  \ln \det \left[1+  \frac{\epsilon^2\tau}{\hbar}\int_0^t dt \,\vec{R}^{T}(t',0)\vec{R}(t',0)\right].
\label{eq:coco0}
\end{equation}

{ As a final step, we can take, inversely, the steps which bring us from Eq.~\eqref{eq:cl1} to~\eqref{eq:cl2}, employing the symplecticity of $\vec{R}$. We multiply Eq.~\eqref{eq:coco0} from left and right with $\vec{R}(t,0)\Omega^{-1}$ and $\Omega\vec{R}^T(t,0)$, respectively, to obtain}:
\begin{equation}
S_2(t)-S_2(t=0) = \ln \Tr \left\{\rho^2(t)\right\}=\frac{1}{2}  \ln \det \left[{\cal{B}}(t,0)+  \frac{\epsilon^2\tau}{\hbar}{\cal{A}}(t)\right].
\label{coco}
\end{equation}
This is to be compared with the classical expression (\ref{Renyi2}), where the quantum initial state playing the role of the initial packet, with $a^2 \leftrightarrow  \hbar$.

If we now let $\hbar \rightarrow 0$ {\em before $\epsilon \rightarrow 0$},  we are in the same situation as the limit of small  size $a$ of the initial packet in the classical case.
If, on the contrary, we make $\hbar \rightarrow 0$ {\em after $\epsilon \rightarrow 0$} the result is not the Lyapunov expansion. 

\section{Quantum Pesin relations}

 \label{sec:PesinQ}

Several quantum generalizations of the KS entropy were suggested previously~\cite{Connes_etal1987,Alicki&Fannes1994,Gharibyan2019,Hallam2019}.  In particular, Ref.~\cite{Gharibyan2019} defined an entropy via Quantum Lyapunov exponents which are in turn defined by the Out-of-Time-Order Commutator (OTOC)~\cite{Larkin&Ovchinnikov1969}: the square of a commutator of two operators at different times, $[\mathcal{O}^{\alpha}(t),\mathcal{O}^{\beta}(0)]^2$. We  now make a connection between this notion and our formula~\eqref{eq:Gen1}.

 Let us split { ${\bf A}$ into early and late times as follows:
 \begin{eqnarray}
{\bf A^-} &\equiv&  \sum_{ \alpha }\int_0^{\bar t} dt'\,
\begin{pmatrix}
\bra{\phi_0}x_c^{\alpha}(t') \\
\bra{\phi_0}x_c^{\alpha*}(t')
\end{pmatrix}
\begin{pmatrix}
\bra{\phi_0}x_c^{\alpha}(t') 
& \bra{\phi_0}x_c^{\alpha*}(t')
\end{pmatrix}, \nonumber \\
{\bf A^+} &\equiv&  \sum_{ \alpha }\int_{\bar t}^{ t} dt'\,
\begin{pmatrix}
\bra{\phi_0}x_c^{\alpha}(t') \\
\bra{\phi_0}x_c^{\alpha*}(t')
\end{pmatrix}
\begin{pmatrix}
\bra{\phi_0}x_c^{\alpha}(t') 
& \bra{\phi_0}x_c^{\alpha*}(t')
\end{pmatrix},
\end{eqnarray}
such that:
\begin{equation}
 t \; h_{KS}^{(2)}  = \frac 12 \ln  \det\left[1+\frac{\epsilon^2\tau}{\hbar^2}{\bf A^+} + \frac{\epsilon^2\tau}{\hbar^2}{\bf A^-} \right].
\label{eq:Gen010}  
\end{equation} 
}
Now, because both ${\bf A^{+}}+1$ and ${\bf A}^{-}$ are positive semi-definite, we may apply Minkowski's determinant inequality:
\begin{equation}
  \det\left[1+\frac{\epsilon^2\tau}{\hbar^2}{\bf A^+} + \frac{\epsilon^2\tau}{\hbar^2}{\bf A^-} \right]^{1/2L} \ge   \det\left[ 1 +\frac{\epsilon^2\tau}{\hbar^2}{\bf A^+}  \right]^{1/2L} + \det\left[ \frac{\epsilon^2\tau}{\hbar^2}{\bf A^-} \right]^{1/2L},
\label{eq:aa}  
\end{equation} 
which gives
\begin{equation}
\ln  \det\left[1+\frac{\epsilon^2\tau}{\hbar^2}{\bf A^+} + \frac{\epsilon^2\tau}{\hbar^2}{\bf A^-} \right] \ge \ln  \det\left[ 1 +\frac{\epsilon^2\tau}{\hbar^2}{\bf A^+}  \right].
\label{eq:cc}  
\end{equation} 

Let us use this inequality to find a lower bound $\bar h_{KS}^{(2)} \le h_{KS}^{(2)} $, which we shall later argue may be exact in some cases. We proceed as follows: we choose the time $\bar t$ such that for $\bar t <t' <t$ we may consider that ${\bf A^+} (t') \sim  {\bf A^+} (t)$. The time $\bar \tau\equiv t-\bar{t}$ is of the order of the autocorrelation of the $x^\alpha$. Then, we define $\bar h_{KS}^{(2)}$ with 
\begin{equation}
 {\bar h^{(2)}_{KS} t} \sim -\ln \det\left| 1  +\frac{\bar \tau \epsilon^2\tau}{\hbar^2}
\sum_{ \alpha }
\begin{pmatrix}
x_c^{\alpha}(t)\ket{\phi_0}   \\
x_c^{\alpha*}(t)\ket{\phi_0}
\end{pmatrix}
\otimes
\begin{pmatrix}
\bra{\phi_0} x_c^{\alpha}(t)  \\
\bra{\phi_0} x_c^{\alpha*}(t)
\end{pmatrix} \right|^{-1/2}.
\label{eq:Gen41}
\end{equation}
We can then use the Matrix Determinant Lemma (Eq.~\eqref{eq:MDL}) to obtain, under these assumptions, an expression {\em  in terms of a  $2N\times 2N$ determinant}: 
$
 {\bar h^{(2)}_{KS} t}  =
\frac 12 \ln \det\left| \delta_{\alpha\beta} +\frac{\bar \tau \epsilon^2\tau}{\hbar^2} \langle \phi_0 |x_c^{\alpha}(t) x_c^{\beta}(t) +x_c^{\alpha*}(t) x_c^{\beta*}(t) | \phi_0 \rangle \right|
$, which gives 
\begin{equation}\boxed{
  {\bar h^{(2)}_{KS} t}  =\frac 12 \ln \det\left| \delta_{\alpha\beta} +\frac{\bar \tau \epsilon^2\tau}{\hbar^2} \langle \phi_0 |x_c^{\alpha}(t) x_c^{\beta}(t) +x_c^{\beta}(t) x_c^{\alpha}(t) | \phi_0 \rangle \right|  =\frac 12 \ln\det\left| \delta_{\alpha\beta} +\frac{\bar \tau \epsilon^2\tau}{\hbar^2} \langle \phi_0 |[x^{\alpha}(t),\rho_0] [x^{\beta}(t),\rho_0] | \phi_0 \rangle \right| }
\label{eq:Gen4}
\end{equation}
where we have used the relation in Eq.~(\ref{commu}), the Hermiticity of the $x^{\alpha}$, and have assumed that the Hilbert space is given by real eigenfunctions. Eq.~\eqref{eq:Gen4} is in line with previous works which showed a relation between the second R{\'e}nyi entropy and the OTOC, where the operator at the initial time appearing in the commutator can be taken as the density matrix~\cite{Fan2017,Hosur2016,Bergamasco2019}.

Now the determinants which appear in  {  Eq.~\eqref{eq:Gen4} are} of $2N\times 2N$ matrices.
If there are exponential expansions  and contractions of the $2N\times 2N$ matrix 
\begin{equation}
 {\bf{\hat A}}_{\alpha\beta} \equiv \langle \phi_0 |x_c^{\alpha}(t) x_c^{\beta}(t) +x_c^{\beta}(t) x_c^{\alpha}(t) | \phi_0 \rangle=\langle \phi_0 |[x^{\alpha}(t),\rho_0] [x^{\beta}(t),\rho_0] | \phi_0 \rangle,
 \label{lyapu}
\end{equation}
 with eigenvalues $e^{\lambda_i t}$, then  thanks to the $\delta_{\alpha\beta}$ term we have 
$$
\bar h_{KS}^{(2)} = \sum_{\lambda_i>0}\lambda_i.
$$
Note the somewhat optimistic  notation: we have not shown that the $\lambda_i$ deserve to be called individually `Lyapunov exponents', although it is very tempting to interpret them this way.  Indeed, if there are exponential expansions, we may expect the inequality to be saturated, and $\bar h^{(2)}_{KS} = h^{(2)}_{KS}$: this is because the smallness of the time-interval $\bar{\tau}$ is
compensated by the exponential expansion. We shall check this in the semiclassical limit.

A similar relation of the form $h_{KS}^{(2)} = \sum_{\lambda_i>0}\lambda_i$ was already proposed in Ref.~\cite{Gharibyan2019}  as a definition of the quantum KS entropy, where the $\lambda_i$ defined instead on the basis of the eigenvalues of the matrix  $\vec{L}^{\alpha\beta}(t)=-\bra{\psi}[x^{\alpha}(t),x^{\gamma}(0)][x^{\gamma}(t),x^{\beta}(0)]\ket{\psi}$, with $\ket{\psi}$ being an eigenstate of the Hamiltonian. (Another different definition  of Lyapunov spectrum was given in Ref.~\cite{Rozenbaum_etal2019}). Let us insist that the relation we derive here is not a definition, but rather a derivation in terms of the entropy production.\\

{\bf Check of the semiclassical limit}\\

{ One can check the semiclassical limit on second equality of Eq.~\eqref{eq:Gen4}: 
As shown in Sec.~\ref{sec:SClassical}, we have $x^{\alpha}_c(t)\xrightarrow[\hbar\to 0]{}R^{\alpha\gamma}(t,0)y^{\gamma}(0)$, thus
$$
\bra{\phi_0}x_c^{\alpha}(t)x_c^{\beta}(t)+x_c^{\beta}(t)x_c^{\alpha}(t)\ket{\phi_0}
=\sum_{\gamma\delta}R^{\alpha\gamma}(t,0)R^{\beta\delta}(t,0)\bra{\phi_0}y^{\gamma}(0)y^{\delta}(0)+y^{\delta}(0)y^{\gamma}(0)\ket{\phi_0}.
$$
The matrix $\bra{\phi_0}y^{\gamma}(0)y^{\delta}(0)+y^{\delta}(0)y^{\gamma}(0)\ket{\phi_0}$ is simply the identity matrix times $\hbar$: recall that for the semiclassical limit we choose $\ket{\phi_0}$ as the oscillator vacuum, therefore, if the indices correspond to different degrees of freedom, the 
expectation breaks into a product of two expectations both vanishing, if they correspond to a $p^a$ and its partner $q^a$ they also vanish, and $\bra{\phi_0}[y^{\gamma}(0)]^2\ket{\phi_0}=\hbar$. Then:
\begin{equation}
\bar h^{(2)}_{KS} t \xrightarrow[\hbar\to 0]{} \ln \det\left| 1 +\frac{\bar \tau \epsilon^2\tau}{\hbar} R(t,0)R^T(t,0) \right|^{-1/2}=
\frac 1 2 \det\left| 1 +\frac{\bar \tau \epsilon^2\tau}{\hbar} {\cal{B}}(t,0) \right|,
\end{equation}
}
which is the correct classical result, and thus $\bar h^{(2)}_{KS}=h^{(2)}_{KS}$ in this case.

\section{Averaging over initial states}
\label{sec:IC}

{ In the analysis above we have assumed that: ({\it i}\,) the Hilbert space is represented by real and orthogonal eigenfunctions $\ket{\mu}$, and ({\it ii}\,) that the initial state is given by one of these eigenfunctions $\rho_0=\ket{\phi_0}\bra{\phi_0}$.  These assumptions facilitate the derivation of the KS entropy in Eq.~\eqref{eq:Gen00}, and were suitable for taking its semiclassical limit. We now discuss how to generalize the result beyond these assumptions.

Concerning point ({\it i}\,), it is important to extend the result to a non-real eigenbasis since the derivation, as well as the proof of additivity, assumes a basis which is orthonormal to $\ket{\phi_0}$. We can always construct such a basis, however, depending on $\ket{\phi_0}$, it might not be real. The derivation can be readily extended to any choice of eigenbasis $\{\ket{\mu}\}$. The assumption of real eigenbasis was only to facilitate the notations for $\left(\bra{\phi_0}x^{\alpha}(t')\right)^*=\bra{\phi_0}x^{\alpha *}(t')$. For the general case, the matrix $\bf A$, which is constructed by four blocks, reads
$$
\vec{A}
=
\sum_{ \alpha }\int_0^tdt'\,
\begin{pmatrix}
\bra{\mu}x_c^{\alpha}(t')\ket{\phi_0}  \bra{\phi_0}    x_c^{\alpha}(t')\ket{\nu}   \qquad   \bra{\mu}x_c^{\alpha}(t')\ket{\phi_0}   \bra{\phi_0}  x_c^{\alpha}(t')\ket{\nu}^* \\
\bra{\mu}x_c^{\alpha}(t')\ket{\phi_0}^*  \bra{\phi_0}    x_c^{\alpha}(t')\ket{\nu}   \qquad     \bra{\mu}x_c^{\alpha}(t')\ket{\phi_0}^*  \bra{\phi_0}    x_c^{\alpha}(t')\ket{\nu}^*
\end{pmatrix}.
$$
}

{ We are left with relaxing assumption ({\it ii}): let us now show how we can average over initial conditions, generalizing the result beyond the case of $\rho_0=\ket{\phi_0}\bra{\phi_0}$.  The state $|\phi_0\rangle$} may be chosen as the ground state of an oscillator, translated by $(q_o^i,p_o^i)$. In other words, $|\phi_0\rangle$ is the coherent state
$|z\rangle$:
\begin{equation}
|\phi_0 \rangle = e^{i q^i_o p_i - i p^i_o q_i} |0\rangle = e^{z_i \alpha^\dag_i-z^*_i \alpha_i}  |0\rangle =  |\vec z\rangle
\end{equation}
where $z^a=q^a_o+i p^a_o$ and $\alpha^\dag_a= q_a-i p_a$.

This immediately suggests how to average over initial conditions. Consider for example the Boltzmann-Gibbs measure $\rho_{GB}$. Like any operator,
it may be expressed in its $P$-symbol representation \cite{perelomov}
\begin{equation}
\rho_{GB}= \int dz d\bar z \; \rho_{GB}^P(z,\bar z)   \; |z\rangle\langle z|,
\end{equation} where $\rho_{GB}^P(z,\bar z) $ is a phase-space function representing $\rho_{GB}$.
Then, a R{\'e}nyi entropy averaged over the Gibbs-Boltzmann entropy reads:
\begin{equation}
\left\langle e^{h_{ks}^{(2)} t} \right\rangle_{GB} = - \int dz d\bar z \; \rho_{GB}^P(z,\bar z)   \Tr_z \left\{\rho^2(t)\right\},
 \end{equation}
 and 
\begin{equation}
 \Tr_z \left\{\rho^2(t)\right\}=\det \left|1+\frac{\epsilon^2\tau}{2\hbar^2}\vec{A}_z\right|^{-1/2},
\label{eq:Gen8}
\end{equation}
with ${\bf A_z}$ defined as ${\bf A}$ in Eq.~\eqref{eq:Gen1}, replacing $\ket{\phi_0}$ with $\ket{z}$.
The same can obviously be done  for all R{\'e}nyi entropies $S_q$.
Note that it is in this averaging over $z$ that the nontrivial dependence on $q$ (`multifractality') enters.

\subsection{Other groups, localized initial state}

Writing the R{\'e}nyi entropy with averaging over coherent states, the generalization to spin systems seems natural. Suppose we have a spin-$j$ chain. We construct the coherent states for each site
in this representation in the standard way: 
\begin{equation}
|z \rangle =\Pi_i |z_i\rangle =e^{\sum_i z_i J^+_i}|0\rangle \qquad ; \qquad J^-_i|0\rangle=0.
\end{equation}
 Next, we consider two independent operators per site, for example $J_i^x$ and
$J_i^y$, and construct in this way the corresponding operator $A = \int dt' \sum_i^\alpha  J_i^\alpha(t') |z\rangle \langle z|J_i^\alpha(t')$, where $\alpha=x,y$.
All the previous steps follow, including the semiclassical limit for $j \rightarrow \infty$.

For fermion systems, a similar strategy may be envisaged, replacing the spin operators by bi-fermionic operators $a^\dag_i a_j$.

\section{Summary and outlook}
\label{sec:Dis}

We have discussed a Kolmogorov-Sinai entropy of a closed quantum system by coupling it weakly to an auxiliary bath. 
The construction here allows one to derive (rather than assume)  a quantum version of the Pesin relation.
The external bath is classical in the sense that it is Markovian. However, the same steps of the analysis may be generalized to a truly quantum bath with a finite temperature and friction through the Schwinger-Keldysh formalism. { We note that adding noise correlation can simply be done through the matrix $\vec{K}$ in Eq.~\eqref{eq:UV}. However, the case of friction is less trivial, since within the Schwinger-Keldysh formalism, it means that forward and backward trajectories are not identical, and the second Renyi entropy in Eq.~\eqref{eq:Trho2} will thus involve four realizations of the noise.}

In this paper we considered the instability generated in the system by the bath when coupled to a set of operators that is of the order of
the number of degrees of freedom, or a subset of them. This has the important consequence that some final expressions
may be written in lower-dimensional spaces.  One could  of course  have used a much larger set of operators, including projectors onto pure states, and  the result would have probably been different. Our procedure is thus an arbitrary, if physically motivated, choice.

A possibility that is physically interesting is to  couple noise to a set of operators that are local in real space.
As noted above, definition~\eqref{eq:Gen00} is not restricted to systems where there is an exponential growth of OTOC-like quantities, and some, though not all, of the results in this paper carry through when there are no exponential growths in time, as is believed to be the case in local  spin systems.

It would be also interesting to establish contact with the mathematical literature on the subject, in particular with the  Connes-Narnhofer-Thirring entropy~\cite{Connes_etal1987}. Note that this entropy is defined in the thermodynamic limit, thus, connection between the two definitions shall be done in view of the macroscopic limit.

The recent bound on the quantum Lyapunov exponent states that the exponential growth rate of the OTOC cannot be larger than $2\pi k_B T/\hbar$, with $T$ the temperature~\cite{Maldacena_etal2016}. This result has simulated an enormous body of  theoretical and experimental studies in various fields. 
An interesting future direction will be to study a quantum bound on the KS entropy. Indeed, it is natural to have a temperature-dependent bound on the growth rate of an entropy, which, unlike the Lyapunov exponent,  is an extensive thermodynamic quantity. 
Finally, it will be interesting to understand whether the production rate of entanglement entropy in a closed quantum system, which might be driving thermalization~\cite{Kaufman2016}, follows the KS entropy. Several works have already established such a relation in semiclassical setups~\cite{Miller&Sarkar1999,Monteoliva&Paz2000,Bianchi_etal2015,Asplund&Berenstein2016} or by projecting the quantum unitary evolution to an effective semiclassical one~\cite{Hallam2019}.

\begin{acknowledgments}
TG and JK are supported by the Simons Foundation Grant No. 454943.
\end{acknowledgments}

\onecolumngrid

\appendix

\section{Classical derivation}
\label{app:classical}

In this appendix we provide ({\it i\,}) rigorous grounds for the classical Pesin relation given in Sec.~\ref{sec:cl}, ({\it ii\,}) a classical derivation involving Poisson brackets, which is completely analog to the quantum one, and ({\it iii\,}) an alternative semiclassical derivation.

\subsection{Classical Pesin relation: more details}
\label{app:susy}

In order to see how contributions from the noise at different times add, the natural objects to consider are the $p$  forms $\Lambda_{i_1,...,i_p}(t)$, which represent the volume elements of dimension
$p$. These expand according to \cite{Paladin&Vulpiani1986}  
$\Sigma_{i_1,...,i_p} \Lambda_{i_1,...,i_p}^2(t) \sim    e^{t \Lambda^{(p)}} \; \Sigma_{i_1,...,i_p} \Lambda_{i_1,...,i_p}^2(0) \sim e^{(\lambda_1+...+\lambda_p)t} \; \Sigma_{i_1,...,i_p} \Lambda_{i_1,...,i_p}^2(0)$. 
For a physicist, the easiest way to represent this is introducing fermions $a_i,a^\dag_i$ and their vacuum state $|0\rangle$, and writing the contribution from all the $t'$ as~\cite{Gozzi&Reuter1994,Tanase_Nicola&Kurchan2003}:
\begin{equation}
\Lambda_{i_1,...,i_p}(t) = \int dt' \; \langle 0| a_{i_1} ... a_{i_p}  (R_{i_il_1}(t,t') a^\dag_{l_1})  ...  (R_{i_p l_p}(t,t') a^\dag_{l_p}) |0\rangle  
\end{equation}
In fact, introducing another family of fermions $b_i, b^\dag_i$, and the vacuum of both species, we have 
\begin{eqnarray}
& &\Sigma_{i_1,...,i_p} \Lambda_{i_1,...,i_p}^2(t) =  \Sigma_{i_1,...,i_p} \int dt' \; \nonumber \\ & &
\langle 0|   a_{i_1} ... a_{i_p} ... b_{i_1} ... b_{i_p} (R_{i_im_1}(t,t') a^\dag_{m_1})  ...  (R_{i_p m_p}(t,t') a^\dag_{m_p})
 (R_{i_il_1}(t,t') b^\dag_{l_1})  ...  (R_{i_p l_p}(t,t') b^\dag_{l_p}) |0\rangle \nonumber\\
 &=& \Sigma_{i_1,...,i_p} \int dt' \; \langle 0|   a_{i_1} ... a_{i_p} ... b_{i_1} ... b_{i_p} \;\; \left\{    {\cal T} e^{- \int_{t'}^t\hat H(t'')  dt''}   \right\}    \;   \;        b^\dag_{i_1} ... b^\dag_{i_p} ... a^\dag_{i_1} ... a^\dag_{i_p} |0\rangle 
\end{eqnarray}
where $\hat H =\sum_{\alpha \gamma} 
\Omega^{\alpha \beta}  \frac{\partial^2 {\cal H}}{\partial x_\beta \partial x_\gamma} (a^\dag_{\alpha} a_\gamma + b^\dag_{\alpha} b_\gamma)$ generates the linear
 transformation. In particular
\begin{eqnarray}
& &\det A^2(t) =  \Sigma_{i_1,...,i_{2N}} \int dt' \; \langle 0|   a_{i_1} ... a_{i_{2N}} ... b_{i_1} ... b_{i_{2N}} \;\;  \left\{   {\cal T} e^{- \int_{t'}^t\hat H(t'') dt''  }   \right\}   \;   \;        b^\dag_{i_1} ... b^\dag_{i_{2N}} ... a^\dag_{i_1} ... a^\dag_{i_{2N}} |0\rangle 
\end{eqnarray}
Now, the existence of Lyapunov exponents is the statement that the integral $  {\cal T} e^{- \int_{t'}^t\hat H(t'') dt''  }  \sim e^{(t-t') \hat H_o} $ is extensive 
to exponential accuracy in time, so that  $  {\cal T} e^{- \int_{t'}^t\hat H(t'') dt''  }  \sim e^{(t-t') \Lambda^{(p)}} |p \rangle \langle p|$, where $ \Lambda^{(p)}$ and $|r \rangle$
are the lowest eigenvalue and corresponding eigenvector of the subspace with $p$ fermions of each kind. 

We may thus evaluate the integrals to within exponential
accuracy to get
\begin{eqnarray}
& &\det \mathcal{A}^2(t) \sim  \Sigma_{i_1,...,i_{2N}} \; \langle 0|   a_{i_1} ... a_{i_{2N}} ... b_{i_1} ... b_{i_{2N}} \;\;   \left(e^{- t \hat H_o   } +1 \right) \hat H_o^{-1}    \;   \;        b^\dag_{i_1} ... b^\dag_{i_{2N}} ... a^\dag_{i_1} ... a^\dag_{i_{2N}} |0\rangle 
\end{eqnarray}
The identity added in the central bracket comes precisely from the times $t' \sim t$ and encompasses the effect of the late noise on the contracting modes. We may now
distribute the expression for the determinant, which becomes a development in minors:
\begin{equation}
\det \mathcal{A}^2(t) \sim \Sigma_p    c_p e^{t  \Lambda^{(p)}}  [ \Lambda^{(p)}]^{-1}  \propto e^{t  \Lambda^{(p^+)}}  
\end{equation}
where the $c_p$ are time-independent overlaps.
The sum is dominated by the exponential of largest of the all the exponents, say $\Lambda^{(p^+)}$, which we hence identify as the KS entropy.
Similarly, the expansion of $p$ forms, with $p \le p^+$  is dominated by $\Lambda^{(p)}$, so we may identify $\Lambda^{(p)}$ as the sum of the first $p$ Lyapunov exponents.
Then, $h_{KS}= \Lambda^{(p^+)}$ which is precisely the sum of the positive Lyapunov exponents.  For a dynamics averaged over a strong noise, it is easy to make this into a rigorous proof.

\subsection{Classical formula with Poisson brackets: direct analogy with the quantum case.}
\label{app:Poisson}

We now derive a classical formula which is an analog of Eq.~\eqref{eq:Gen0}. We show how it yields the classical result given in Eq.~\eqref{Renyi2}.
The way to go from classical to quantum mechanics is 
\begin{equation}
-i[\cdot,\cdot]\rightarrow-\{\cdot,\cdot\},\qquad \Tr\rightarrow \int dx.
\end{equation}
Thus, we expect that for classical Hamiltonian system Eq.~\eqref{eq:Gen0} transforms as
\begin{multline}
e^{-h_{KS}^{(2)}t } \propto \int\mathcal{D}{\bm \eta}P[{\bm \eta}]
 \exp\left(
-\frac{1}{2}\sum_{\alpha,\beta}\int dt'\,dt''\,\int dx_0\frac{\left\{i\left\{x^{\alpha}(t'),\rho_{0}(x_0)\right\}i\left\{x^{\beta}(t''),\rho_{0}(x_0)\right\}\right\}}{ \int dx_0 \rho^2_0(x_0)}\eta^{\alpha}(t') \eta^{\beta}(t'')\right).
\label{eq:Gen0Cl}
\end{multline}
Let us now see this in detail.

Classically, we have $\frac{d x}{d t}= -i\mathcal{L}x$, and for the density matrix
$\frac{d \rho}{d t}= i\mathcal{L}\rho$, where $ -i\mathcal{L}\equiv -\left\{\cdot,H\right\}$.
The $\mathcal{L}$ superoperator is Hermitian. We can define the following superoperator
\begin{equation}
\mathcal{U}(t,t') f\equiv\mathcal{T}e^{-i\int^t_{t'}\mathcal{L}(t'')dt''}f=\mathcal{T}e^{-\int^t_{t'}\left\{\cdot,H(t'')\right\}dt''}f.
\end{equation}
Now, let us turn  to our problem. We have the following time-dependent Hamiltonian
\begin{equation}
H_h=H_0+h(t)x.
\end{equation}
we wish to calculate the R{\'e}nyi entropy at time $t$, starting with some initial distribution of points in phase-space $\rho_0$.  Thus we look at
\begin{equation}
\int dx_t \rho^2(x_t)=\int \mathcal{D}h_1 \mathcal{D}h_2  e^{\ln\mathcal{Z}(h_1,h_2)}, 
\end{equation}
with 
\begin{equation}
\mathcal{Z}(h_1,h_2)\equiv\int dx_t\left\{\left(\mathcal{U}_{h_1}^{-1}(t,0)\rho_0\right)\left(\mathcal{U}_{h_2}^{-1}(t,0)\rho_0\right)\right\}.
\end{equation}
Expanding to second order around $h_1,h_2=0$ is easier if we write
\begin{equation}
\mathcal{U}_{h_1}^{-1}(t,0)\rho_0=\mathcal{U}_{h_1}^{-1}(t,t'')\mathcal{U}_{h_1}^{-1}(t'',t')\mathcal{U}_{h_1}^{-1}(t',0)\rho_0.
\end{equation}
Then, for example, the term which is proportional to $h_1(t')h_1(t'')$ goes as
\begin{multline}
a_{1,t';1,t'}\propto \int dx_t e^{i\mathcal{L}_0(t-t'')}\left\{e^{i\mathcal{L}_0(t''-t')}\left\{e^{i\mathcal{L}_0t'}\rho_0,x_0\right\},x_0\right\}e^{i\mathcal{L}_0t}\rho_0=\\
\int dx_0 e^{-i\mathcal{L}_0t''}\left\{e^{i\mathcal{L}_0(t''-t')}\left\{e^{i\mathcal{L}_0t'}\rho_0,x_0\right\},x_0\right\}\rho_0=
\int dx_0 \left\{\left\{\rho_0,e^{-i\mathcal{L}_0t'}x_0\right\},e^{-i\mathcal{L}_0t''}x_0\right\}\rho_0=\\
\int dx_0 \left\{\rho_0,e^{-i\mathcal{L}_0t'}x_0\right\}\left\{\rho_0,e^{-i\mathcal{L}_0t''}x_0\right\}
=\int dx_0 \left\{\rho_0,x(t')\right\}\left\{\rho_0,x(t'')\right\}.
\end{multline}
For the first equality we use the identity 
$\int dy f(y) e^{i\mathcal{L}t} g(y) = \int d\tilde{y} e^{-i\mathcal{L}t}f(\tilde{y}) g(\tilde{y})$,
where we unitarily change basis $\tilde{y}\rightarrow e^{i\mathcal{L}t}y$;
for the second equality we used the identity: 
$\mathcal{U}\left\{f,g\right\}  = \left\{\mathcal{U}f,\mathcal{U}g\right\}$; and for the third equality the identity:
$\int \left\{f,g\right\}k = -\int f\left\{k,g\right\}$.
Therefore, in analogy to the quantum case we have
\begin{equation}
a_{1,t';1,t'}\propto 
\int dt' dt''\int dx_0 \{\rho_0(x_0),x(t')\}\{\rho_0(x_0),x(t'')\},
\end{equation}
and one can conclude that the classical KS entropy follows Eq.~\eqref{eq:Gen0Cl}. Next, let us show how this alternative classical formula relates to the one derived in the text, Eq.~\eqref{eq:Gen0}. 

Let us assume some Gaussian initial condition 
\begin{equation}
\rho_0(x_0)=\frac{1}{(2\pi a)^{d/2} }e^{-\frac{x_0^2}{2a}}.
\end{equation}
Plugging this into the argument of the exponent in Eq.~\eqref{eq:Gen0Cl} one finds
\begin{equation}
\int dx_0 \frac{\left\{x^{\alpha}(t'),\rho_{0}(x_0)\right\}\left\{x^{\beta}(t''),\rho_{0}(x_0)\right\}}{\int dx_0 \rho^2_0(x_0)}
=\int dx_0 \frac{\partial x^{\alpha}(t')}{\partial x_0^{\gamma}}
\frac{\partial x^{\beta}(t'')}{\partial x_0^{\gamma'}} \Omega^{\gamma\beta}\Omega^{\gamma'\beta'}\frac{x_0^{\beta} x_0^{\beta'}\rho_0^2}{a^2\int dx_0 \rho^2_0(x_0)} \approx 
 \frac{1}{a} \frac{\partial x^{\alpha}(t')}{\partial x_0^{\gamma}}
\frac{\partial x^{\beta}(t'')}{\partial x_0^{\gamma}},
\end{equation}
where we assumed that $\frac{x_0^{\beta}x_0^{\beta'}\rho_0^2}{\int\rho_0^2} \sim a \delta(x_0^{\beta}-x_0^{\beta'})$.
Eq.~\eqref{eq:Gen0Cl} then reads 
\begin{equation}
\lim_{\hbar\to 0} \Tr \left\{\rho^2(t)\right\}=(4\pi a)^{-d/2}\int\mathcal{D}{\bm \eta}P[{\bm \eta}] 
\exp\left(
-\frac{1}{2}\sum_{\alpha,\beta}\int dt'\,dt''\,  \frac{R^{\alpha\gamma}(t',0)R^{\beta\gamma}(t'',0)}{a} \eta^{\alpha}(t') \eta^{\beta}(t'')\right).
\end{equation}
Finally, to solve the Gaussian integral we can follow the same lines which brought us to Eq.~\eqref{eq:Gen1}, taking the same noise and using the Determinant Lemma, to find
$$
\ln \Tr \left\{\rho^2(t)\right\}=\frac{1}{2}  \ln \det \left[1+  \frac{\epsilon^2\tau}{a}\int dt \,\vec{R}^{T}(t',0)\vec{R}(t',0)\right].
$$
Then one can repeat the steps indicated after Eq.~\eqref{eq:coco0} to see the agreement with the classical result Eq.~\eqref{Renyi2} in Sec.~\ref{sec:cl}.

\subsection{semiclassical limit of formula~\eqref{eq:Gen0}}
\label{app:sc2}

Below we show how the semiclassical limit of Eq.~\eqref{eq:Gen0} (an expression for the quantum KS entropy) gives the classical result Eq.~\eqref{eq:Gen0Cl}.

Let us start with  Eq.~\eqref{eq:Gen0}. The quantum object which appears in this equation is $\Tr\left(i[x^{\alpha}(t'),\rho_{0}]i[x^{\beta}(t''),\rho_{0}]\right)$. We can treat it in the semiclassical limit by introducing the Weyl transform of an operator 
\begin{equation}
\mathcal{W}\left\{A\right\}(q,p)=\int e^{-ipy/\hbar}\bra{q+y/2}A\ket{q-y/2}\,dy,
\end{equation}
which satisfies the two relations: $\Tr\left(A B\right)=\int dq dq\,\mathcal{W}\left\{A\right\}\mathcal{W}\left\{B\right\}
$ and $\mathcal{W}(A)\star \mathcal{W}(B)=\mathcal{W}(AB)$, with $\star$ being the Moyal product. Inserting these to Eq.~\eqref{eq:Gen0} and taking the semiclassical limit, $f\star g \equiv fg + \frac{i\hbar}{2}\left\{f,g\right\} + O(\hbar^2)$, we find
\begin{equation}
\frac{\Tr\left(i[x^{\alpha}(t'),\rho_{0}]i[x^{\beta}(t''),\rho_{0}]\right)}{\hbar^2 \Tr\rho_0^2}=
\frac{\int\left\{x^{\alpha}(t'),\rho_{0}(x_0)\right\} \left\{ x^{\beta}(t''),\rho_{0}(x_0)\right\}  dx_0}{\int\rho_0^2(x_0) \, dx_0},
\label{eq:sc}
\end{equation}
where it is assumed that $\rho_0(x_0)$ and $x^{\alpha}(t)$ are simple functions of $x_0$ such that in the semiclssical limit they are given by their Weyl transform~\cite{Case2008}. Eq.~\eqref{eq:sc} tells us that the semiclassical limit of Eq.~\eqref{eq:Gen0} is the classical expression in Eq.~\eqref{eq:Gen0Cl}.

\section{Quantum Derivation}
\subsection{Perturbation Theory}
\label{app:Taylor}

We now provide more details on the derivation of Eq.~\eqref{eq:Gen0}.
We shall expand $\ln\Tr\rho^2(t)$ in small $\vec{h}_1(t)$, $\vec{h}_2(t)$, for the Hamiltonian $\mathcal{H}=H_0(\vec{x})+\sum_\alpha h^{\alpha}(t)x^{\alpha}$. The density matrix evolves as
$$
\rho(t)=U_{h}(t,0)\rho_0 U_{h}^{\dagger}(t,0),
$$
with $U_{h}(t,0)\equiv \mathcal{T}\exp\left( -\frac{i}{\hbar}\int\sum_{\alpha} h^{\alpha}(t')x^{\alpha}dt' \right)$. This can be simply seen from taking $\rho_0=\ket{\phi_0}\bra{\phi_0}$.

Let us consider the derivatives of $\Tr\rho^2(t)=\Tr\left[U_{h_1}(t,0)\rho_0 U_{h_1}^{\dagger}(t,0)U_{h_2}(t,0)\rho_0 U_{h_2}^{\dagger}(t,0)\right]$. We have that
\begin{equation}
\frac{\partial \Tr\rho^2(t)}{\partial h^{\alpha}_{1}(t')}=\frac{\partial \Tr\rho^2(t)}{\partial h^{\alpha}_{2}(t')}=0.
\end{equation}
The second derivative consists several terms: 
\begin{multline}
\hbar^2\frac{\partial^2 \Tr\rho^2(t)}{\partial h^{\alpha}_{1}(t')\partial h^{\beta}_{1}(t'')}=
\hbar^2\frac{\partial^2 \Tr\rho^2(t)}{\partial h^{\alpha}_{2}(t')\partial h^{\beta}_{2}(t'')}=
-\hbar^2\frac{\partial^2 \Tr\rho^2(t)}{\partial h^{\alpha}_{1}(t')\partial h^{\beta}_{2}(t'')}=
-\hbar^2\frac{\partial^2 \Tr\rho^2(t)}{\partial h^{\alpha}_{2}(t')\partial h^{\beta}_{1}(t'')}=\\
-\Tr\left\{x^{\alpha}(t')x^\beta(t'') \rho_0^2+x^{\alpha}(t')x^\beta(t'') \rho_0^2
-x^{\alpha}(t')\rho_0 x^{\beta}(t'')\rho_0-x^{\beta}(t'')\rho_0 x^{\alpha}(t')\rho_0
\right\}=\Tr\left\{[x^{\alpha}(t'),\rho_0][x^{\beta}(t''),\rho_0]\right\},
\end{multline}
where we define $x(t)\equiv U^{\dagger}_{h=0}(t,0)\vec{x}U_{h=0}(t,0)=e^{\frac{i}{\hbar}H_0(\vec{x})t}\vec{x}e^{-\frac{i}{\hbar}H_0(\vec{x})t}$.

In summary, we find that 
$$
\mathcal{Z}[\vec{h_1},\vec{h_2}]\equiv \ln\Tr \left\{U^{\dagger}_{h_2}U_{h_1}\rho_{0}U^{\dagger}_{h_1}U_{h_2}\rho_{0}\right\}\approx \frac{1}{2}\sum_{\alpha,\beta}\int dt' \, dt''\,\frac{\Tr\left\{[x^{\alpha}(t'),\rho_0][x^{\beta}(t''),\rho_0]\right\}}{\hbar^2\Tr\rho_0^2}(h^{\alpha}_1(t')-h^{\alpha}_2(t'))(h^{\beta}_1(t'')-h^{\beta}_2(t'')),
$$
which gives Eq.~\eqref{eq:Gen0}.

\subsection{ All the R{\'e}nyi entropies}
\label{app:Renyi}

We can generalize the derivation to the R{\'e}nyi entropy of any order. Taking $\rho_0=\ket{\phi_0}\bra{\phi_0}$ we can write 
\begin{equation}
t h^{(q)}_{KS} =  \ln \Tr \left\{\rho^q(t)\right\}=\int\Pi_r \mathcal{D}\vec{h}_rP(\vec{h}_r)e^{\sum^{q}_{r=1}\ln\mathcal{Z}(\vec{h}_{r-1},\vec{h}_{r})},
\end{equation}
with $\mathcal{Z}(\vec{h}_{r-1},\vec{h}_{r})\equiv \bra{\phi_0}U^T_{h_{r-1},I}U_{h_{r},I}\ket{\phi_0}$, and $\vec{h}_{0}=\vec{h}_{q}$. Expanding in Taylor series 
\begin{equation}
\sum^{q}_{r=1}\ln\mathcal{Z}(\vec{h}_{r-1}^\alpha,\vec{h}_{r}) \sim \frac 12 \sum^{q}_{r=1}\; \sum_{\alpha,\beta} \frac{\partial^2\ln\mathcal{Z}}{\partial {h}_{r-1}^\alpha \partial {h}_{r}^\beta}  \;  {h}_{r-1}^\alpha  {h}_{r}^\beta
\end{equation}
The second derivatives do not depend on $r$ (replica symmetry), so that we obtain a tight-binding problem with $r$ sites. This may be treated as the case of $q=2$ discussed in the main text,
and  one easily gets: 
\begin{equation}
\ln \Tr \left\{\rho^q(t)\right\}=-\frac12 \sum_{r=1}^{q-1} \ln \det_{\mu, \nu}  \left|\delta^{\mu \nu}+\frac{\sin^2\left(\frac{\pi r}{q}\right)\epsilon^2\tau}{2\hbar^2}A^{\mu \nu}\right|-\ln q+(q-1)\ln 2,
\label{eq:Genq}
\end{equation}
where the factors $\sin^2\left(\frac{\pi r}{q}\right)$ are the eigenvalues of the matrix with ones in the diagonal and minus one-half between nearest-neighbors, and periodic boundary
conditions.
Note that we have not yet averaged over different initial conditions $\rho_0$.

 All the determinants differ only in a  prefactor multiplying $A^{\mu \nu}$, so we conclude that, at least for $q$ of order one, there is no multifractality of quantum origin, i.e. given {\em one} initial quantum condition. But of course there will be once we decide to add over different initial conditions, just as there is in the classical case.

~\\~

\bibliography{KSentropy}

\end{document}